\documentclass[12pt]{article}
\usepackage{epsf}
\usepackage{latexsym}
\def\be{\begin{equation}}
\def\ee{\end{equation}}
\def\bea{\begin{eqnarray}}
\def\eea{\end{eqnarray}}

\begin{document}
\begin{titlepage}
\begin{center}
{\Large \bf William I. Fine Theoretical Physics Institute \\
University of Minnesota \\}
\end{center}
\vspace{0.2in}
\begin{flushright}
FTPI-MINN-13/40 \\
UMN-TH-3312/13 \\
November 2013 \\
\end{flushright}
\vspace{0.3in}
\begin{center}
{\Large \bf Remarks on double Higgs boson production by gluon fusion at threshold
\\}
\vspace{0.2in}
{\bf Xin Li$^a$  and M.B. Voloshin$^{a,b,c}$  \\ }
$^a$School of Physics and Astronomy, University of Minnesota, Minneapolis, MN 55455, USA \\
$^b$William I. Fine Theoretical Physics Institute, University of
Minnesota,\\ Minneapolis, MN 55455, USA \\
$^c$Institute of Theoretical and Experimental Physics, Moscow, 117218, Russia
\\[0.2in]

\end{center}

\vspace{0.2in}

\begin{abstract}
The amplitude of double Higgs boson production by the gluon fusion, $gg \to hh$, is known to be small due to cancellation between the graphs with the boson trilinear coupling and those with the coupling to the top quark. For this reason a study of this process was suggested as a sensitive probe of the Higgs sector nonlinearity. We calculate in a closed analytical form this amplitude at the threshold of the two bosons, where the cancellation is the strongest, and discuss the origin of the small value of the amplitude. We also note that the cancellation in the double boson production is in fact a part of a more general phenomenon of suppression of similar threshold amplitudes for multiple boson production, which, although not directly relevant to the actual top quark and the Higgs boson, can be useful in other studies.
\end{abstract}
\end{titlepage}

With the observation~\cite{hc,ha} of what is most likely the long anticipated Higgs boson of the Standard Model, a further study of the Higgs sector becomes a matter of practical feasibility. In particular the nonlinear terms in this sector, describing the interaction between the bosons, are most fundamentally related to the underlying framework of the Standard Model. Thus a test of the self interaction in the Higgs sector would certainly justify overcoming the experimental difficulties that such study inevitably entails. The specific process in which the Higgs trilinear coupling can be measured at a hadron collider is the double boson production by gluon fusion~\cite{psz,cgmppw,hkmt,dkmz,gm}: $gg \to hh$. At the lowest loop level this process is contributed by two types of graphs shown in Fig.~1, the box diagram and the triangle diagram with the trilinear coupling between the bosons. It has been noticed some time ago~\cite{psz} that with the standard couplings the contributions of these two types of graphs exactly cancel in the limit where the mass $m$ of the top quark is much larger than any kinematical invariant in the process, which also implies that $m \gg \mu$ with $\mu$ being the mass of the $h$ boson. With the actual masses, $m \approx 173\,$GeV and $\mu \approx 126\,$GeV the cancellation is not complete, but still the cross section calculated~\cite{cgmppw,hkmt} with the gluon distribution functions at the LHC energies is greatly suppressed in comparison with what would be given by only one type of graphs in Fig.~1. This suppression of the double Higgs boson production by gluon fusion implies an enhanced relative importance of higher loop corrections~\cite{hkmt} and of any nonstandard couplings~\cite{cgmppw,hkmt}, thus providing an advantageous opportunity for studying the latter effects. 

\begin{figure}[ht]
\begin{center}
 \leavevmode
    \epsfxsize=12cm
    \epsfbox{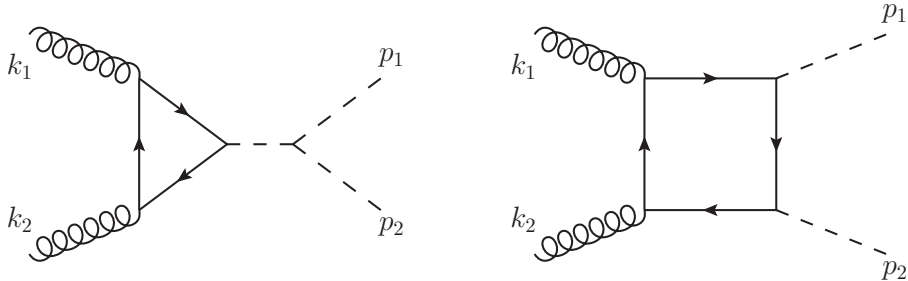}
    \caption{Tho types of diagrams contributing to the amplitude of the process $gg \to hh$: the triangle and the box.}
\end{center}
\end{figure} 

Furthermore, a study of the process $gg \to hh$ may include not only the measurement of the total cross section, but also of the distribution in the invariant $s$ for this process, in particular near the threshold at $s=4\mu^2$, where the effect of the cancellation of the standard contributions is the strongest. It thus appears interesting to analyze in more detail the threshold limit of the amplitude for the the double boson production. It should be noted that a full calculation of both the triangle and the box graphs is available~\cite{psz} at an arbitrary kinematics in terms of the Passarino-Veltman integrals~\cite{pv}, however any actual calculations are so far done by numerical routines, which also somewhat obscures the reasons for the suppression of the process. In this paper we calculate the threshold amplitude in a closed analytical form for arbitrary ratio $\mu/m$ and argue that the `residual' cancellation between the box and triangle graphs at the actual values of $m$ and $\mu$ results from a combination of the exact cancellation in the limit $\mu/m \to 0$, the analytical properties of the amplitude, and the zero of the major absorptive part of the amplitude at $m=\mu/2$ (in addition to the trivial zeros at $m=0$  and $m=\mu$), which can be traced to the property of `nullification'~\cite{mv93} i.e. of exact vanishing of the on-shell sum of the tree level threshold amplitudes for $t \bar t \to n \, h$ at the special mass ratio $m/\mu = N/2$ with integer $N$ and $n \ge N$. The cancellation in the one loop amplitude makes the process sensitive to higher loop corrections. In particular the top quark loop correction to the boson trilinear coupling~\cite{hkmt} produces a singular in the limit $\mu \ll m$ contribution to the amplitude, whose numerical value almost equals that of the one loop term. In the concluding part we also illustrate that the cancellation between different graphs for the double Higgs boson production is in fact a part of more general phenomenon of a similar cancellation in the threshold amplitudes for multiple boson production. Although phenomenologically this behavior is not very significant for the actual top quark and the Higgs boson, it can prove to be relevant in other studies. 

The amplitude for the process $gg \to hh$ at the threshold is described by one form factor $F_2$ and can be written in terms of the momenta $k_1,\,k_2$ and the polarization (and color) amplitudes $\epsilon^a_{1}, \, \epsilon^b_2$ of the gluons as
\be
A(gg \to hh) = -{\alpha_s \over 4 \pi} \, (k_1^\mu \epsilon_1^{a \nu} - k_1^\nu \epsilon_1^{a \mu })(k_{2 \, \mu} \epsilon_{2 \, \nu}^a - k_{2 \, \nu} \epsilon_{2 \, \mu}^a ) \, F_2~,
\label{deff}
\ee
where $\alpha_s$ is the QCD coupling constant.
In the limit $\mu/m \to 0$ the field $h$ can be replaced by a constant and the form factor $F_2$ can be found~\cite{egn,svvz,psz} by considering the top quark loop for the vacuum polarization with the mass $m$ rescaled in the constant background: $m \to m(1+h /v)$, where $v= (G_F \, \sqrt{2})^{-1/2} \approx 246\,$GeV is the Higgs field vacuum expectation value. Proceeding in this way one finds
\be
F_2 = {2 \over 3} \, \langle hh | \log \left ( 1 + {h \over v} \right ) |0 \rangle = {2 \over 3} \, \langle hh | \, {h \over v} - {1 \over 2} \, {h^2 \over v^2} \, |0 \rangle~.
\label{hexp}
\ee
Clearly, the quadratic in $h$ term in the latter expansion corresponds to the contribution of the box graph, while the linear in $h$ term describes the contribution of the triangle diagram with the subsequent `self proliferation' of the bosons. With the Standard Model couplings one readily finds for the two bosons produced at the threshold
\be
\langle hh | \,  {h \over v} \, |0 \rangle = {1 \over 2} \, \langle hh | \, {h^2 \over v^2} \, |0 \rangle ={1 \over v^2}~,
\label{12h}
\ee
and verifies the exact cancellation in Eq.(\ref{hexp}) between the box and the triangle.

For finite masses $\mu$ and $m$ the form factor can be written in terms of a dimensionless function $f$ of the ratio $z=\mu^2/m^2$ as $F_2=f(z)/v^2$. The contribution $f_\triangle$  of the triangle graph to the function $f(z)$ can be readily found by a simple adaptation of the analytical expression~\cite{svvz} for the amplitude for the coupling of the Higgs boson to two photons (or gluons):
\be
f_\triangle = z^{-1}  \, \left [ 1 + (1-z^{-1}) \arcsin^2 ( \sqrt{z}) \right ]~,
\label{ftr}
\ee  
where the branch of the function $\arcsin x$ is defined in such a way that on the upper side of the cut at positive real $x > 1$ it reads as 
\be
\left . \arcsin x  \right |_{\, x > 1} = {\pi \over 2} + {i \over 2} \, \log {1+ \sqrt{1-x^{-2}} \over 1- \sqrt{1-x^{-2}}}~.
\label{defas}
\ee

The contribution $f_\Box$ of the box type graphs to the function $f(z)$ can be found using its analytical and asymptotic properties. Indeed, the function $f(z)$ vanishes in the limit corresponding to zero top quark mass, $|z| \to \infty$, and also at $z \to 0$ due to the discussed above low energy theorem. This function is real at real $z$ in the interval $-4 < z < 1$ and has a right cut at positive $z$ starting from $z=1$ and a left cut at $z < -4$. This implies that the function $f(z)$ can be fully restored from its imaginary part on the cuts. The imaginary part can be found from unitarity (or, equivalently, using the Cutkosky's cutting rules). The corresponding cuts for both the triangle and the box graphs are shown in Fig.~2. The triangle graph has only one cut which results together with the cuts of the box diagram of the type in Fig.~2b in the discontinuity of the function $f(z)$ starting at $z=1$. The cuts of the type shown in Fig.~2c contribute to the discontinuity at positive $z$ starting at $z=4$, while the cuts of the type in Fig.2d give rise to the discontinuity in $f(z)$ at negative $z$ such that $z < -4$. It should be noted that for the purpose of calculation of the box graphs the parameter $\mu^2$ refers to the momentum transferred by a scalar source in the vertex in the graph, $p^2=\mu^2$. Thus setting this momentum space-like $p^2 < 0$, corresponding to the `physical region' for the cut of the type of Fig.~2d does not lead to any inconsistency. It is for this reason that the analytical continuation to negative $z=\mu^2/m^2$ should be considered as to a negative $\mu^2$ while preserving $m^2$  positive, since the parameter $m$ enters the diagrams as dynamical in the propagator of the quark.

\begin{figure}[ht]
\begin{center}
 \leavevmode
    \epsfxsize=12cm
    \epsfbox{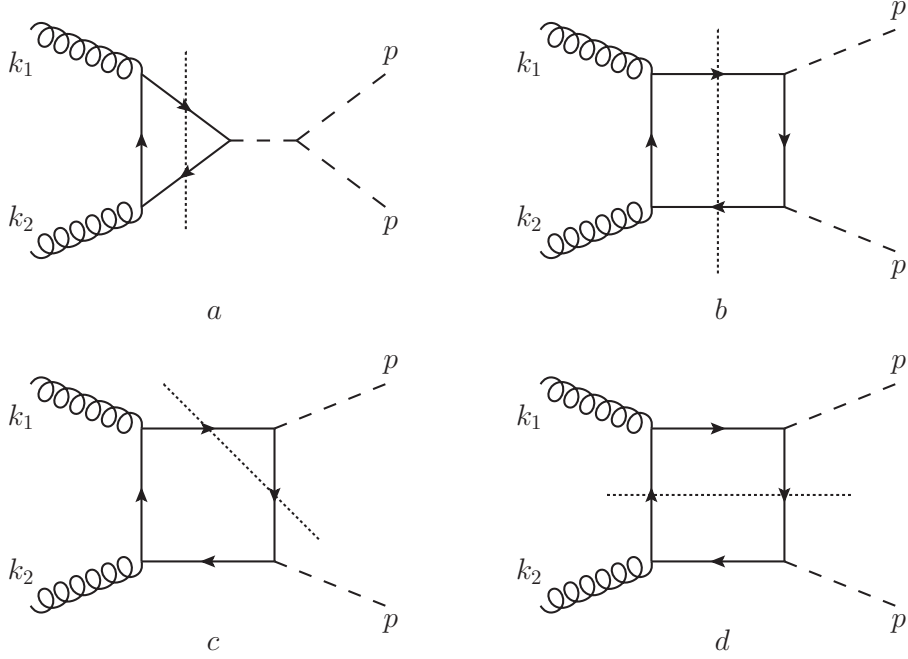}
    \caption{The types of cuts (the dotted lines) describing the imaginary part of the function $f(z)$.}
\end{center}
\end{figure} 

In what follows we denote the imaginary part of the function $f(z)$ resulting from the cuts of each type in Fig.~2 as respectively Im$f_\triangle$, Im$_{(b)}f_\Box$, Im$_{(c)}f_\Box$ and Im$_{(d)}f_\Box$. It is a simple exercise to verify (even before the integration over the phase space of $t \bar t$) that the expressions arising from the cuts in Fig.~2a and 2b are related:
\be
{\rm Im}_{(b)}f_\Box = -{4 m^2 \over \mu^2} \, {\rm Im}f_\triangle~,
\label{relim}
\ee
which implies, given Eq.(\ref{ftr}), that these two expressions combine in the total `$s$-channel' absorptive part Im$_{(s)}f$ of $f(z)$, corresponding to the process $gg \to t \bar t \to hh$ with on-shell quarks, having the form
\be
{\rm Im}_{(s)} f(z) =  {\pi \over 2} \, \theta(z-1) \, z^{-1} \, (1- 4 \, z^{-1}) \, (1- z^{-1}) \, \log {1+ \sqrt{1-z^{-1}} \over 1- \sqrt{1-z^{-1}}}~,
\label{ims}
\ee   
where $\theta$ stands for the step function. One can notice that the expression in Eq.(\ref{ims}) has a nontrivial zero at  $m=\mu/2$ in a complete agreement with the `nullification' property for the amplitudes describing the production of scalars at threshold by on-shell fermions~\cite{mv93}. 

For the imaginary part generated by the cuts of the type in Fig.~2c and Fig.~2d we find after a straightforward calculation
\be
{\rm Im}_{(c)} f = - {\pi \over 2 } \, \theta(z-4) \, z^{-1} \, \left [  \left ( 3 - {4  z^{-1}} \right ) \, \sqrt{1-4 z^{-1}} -  (1+4 z^{-1}) \, (1-2z^{-1}) \,  \log {1+ \sqrt{1-4 z^{-1}} \over 1- \sqrt{1- 4 z^{-1}}} \right ]~,
\label{imc}
\ee  
and
\be
{\rm Im}_{(d)} f =  {\pi \over 2 } \, \theta(-z-4) \, z^{-1} \,  \left [ (3 - 4 z^{-1}) \, \sqrt{1+4 z^{-1}} - (1+4 z^{-1}) \, (1- 2 z^{-1}) \, \log {1+ \sqrt{1+4 z^{-1}} \over 1- \sqrt{1+ 4 z^{-1}}} \right ]~.
\label{imd}
\ee
A strong similarity between the expressions in Eq.(\ref{imc}) and (\ref{imd}) is apparent, but the reason for it is not. 

The full absorptive part of the function $f(z)$ is given by the sum of the expressions (\ref{ims}), (\ref{imc}) and (\ref{imd}). The discontinuity at the cuts and the condition that $f(z)$ goes to zero at $|z| \to \infty$ and that it is also vanishing at $z =0$ is sufficient to restore the full expression for $f(z)$ (e.g. by using the dispersion relation).  The result can be written in a closed analytical form as
\bea
\label{fullf}
&&f(z) =  2 z^{-1}  +
z^{-1} \, (1-4 z^{-1}) \, (1-z^{-1}) \, \arcsin^2 (\sqrt{z}) +  \\
&& z^{-1}  \, \left[ \left ( 3 - 4 z^{-1} \right ) \, \sqrt{4 z^{-1} - 1} \, \arcsin \left ( {\sqrt{z} \over 2} \right ) + (1+4 z^{-1}) \, (1-2z^{-1}) \, \arcsin^2 \left ( {\sqrt{z} \over 2} \right ) \right ] + \nonumber \\
&&z^{-1}  \, \left[ (4 z^{-1} - 3) \, \sqrt{1+4 z^{-1}} \, {\rm arcsinh} \left ( {\sqrt{z} \over 2} \right ) + (1+4 z^{-1}) \, (1-2z^{-1}) \, {\rm arcsinh}^2 \left ( {\sqrt{z} \over 2} \right ) \right ]~. \nonumber
\eea

Numerically, the actual masses of the top quark and the Higgs boson correspond to the value $z=z_0 \approx 0.53$, where the expression (\ref{fullf}) gives $f(z_0) \approx -0.072$. This value is more than ten times smaller and of the opposite sign compared with the contribution of the triangle graph alone: $f_\triangle(z_0) \approx 0.77$. Clearly, such significant cancellation implies that besides the vanishing of $f(z)$ at $z=0$,  the coefficients of the Taylor expansion for $f(z)$ are quite small: 
\be
f(z) = -{7 \over 90} \, z - {1 \over 14} \, z^2 + O(z^3)~.
\label{tf}
\ee
It can be noted that had one ignored the right and left `far' cuts for $f(z)$ starting at $z= \pm 4$ and restored this function from the absorptive part (\ref{ims}) alone (but still using the condition of $f(z)$ vanishing at zero and infinity), the result would be the function 
\be
\tilde f(z) = -4 z^{-2}+ {11 \over 3} \,  z^{-1} - {2 \over 45} \, z^{-1} \, (1-4 z^{-1}) \, (1-z^{-1}) \, \arcsin^2 (\sqrt{z}) = -{31 \over 315} \, z - {12 \over 175} \, z^2 + O(z^3)~,
\label{tilf}
\ee
which gives a reasonably close approximation for $f(z)$ at $|z| < 1$, so that, in this sense, the cut starting at $z=1$ gives the major contribution to the full result in this domain of $z$. 

Using the above formulas one can also estimate, in the limit of small $z$, the behavior of the amplitude slightly above the threshold, namely at the value of $s$ for the two bosons such that $s > 4 \mu^2$, but still $s \ll m^2$. Indeed, the box graph depends on the parameter $s/m^2$ and in this limit can be taken at its threshold value. The only dependence on $s$ then arises from the Higgs boson propagator in the diagram with triangle. The contribution of the triangle graph to the amplitude $gg \to hh$ is still described by one form factor $F_2$ as in Eq.(\ref{deff}) which is given by
\be
F_{2 (\triangle)} = {1 \over v^2} \, f_\triangle \left ( {s \over 4m^2} \right ) \, {3 \mu^2 \over s-\mu^2} = {2 \over 3 v^2} - {2 \over 3 v^2} \, {s-4 \mu^2 \over s- \mu^2} + O \left ( {s \over m^2},\, {\mu^2 \over m^2} \right )~.
\label{f2s}
\ee
The first (constant) term cancels against the box graph contribution, and one can see that the form factor deviates toward negative values above the threshold. This is the same negative sign as the threshold amplitude (\ref{fullf}) at small, but nonzero, values of $z$. Therefore it can be concluded that the discussed cancellation between the box and triangle contributions is the strongest at the threshold.

The small value of the leading Standard Model term for the amplitude of $gg \to hh$ makes it very sensitive, besides possible nonstandard effects, to higher order corrections. At small $z$ the most important correction at the threshold arises from the modification of the Higgs trilinear coupling by a top quark loop as shown in Fig.~3. (This is also the only next to leading order correction enhanced by the number of colors $N_c$.) Indeed, in terms of the function $f(z)$ this correction, $\delta f$, is proportional to $(m^2/v^2) (m^2/\mu^2)$ and is thus singular in $z$ at $z \to 0$. The coefficient of the singularity is recently calculated~\cite{hkmt} by using the effective Higgs potential generated by the top quark loop, and the result reads as
\be
\delta f(z) = - {m^2 \over \pi^2 v^2} \, {f_\triangle(0) \over z} = - {2 m^4 \over 3 \pi^2 v^2 \mu^2} \approx - 0.063~,
\label{deltf}
\ee
which numerical value is only slightly smaller than the leading order result in Eq.(\ref{fullf}). 

\begin{figure}[ht]
\begin{center}
 \leavevmode
    \epsfxsize=8cm
    \epsfbox{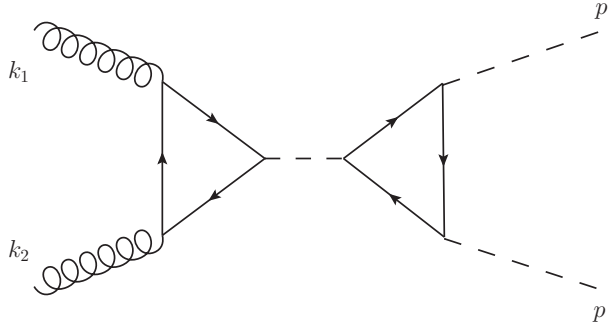}
    \caption{The two loop diagram giving rise to a dominant, singular in the limit $\mu^2/m^2 \to 0$, correction to the form factor $F_2$.}
\end{center}
\end{figure}

Before concluding our discussion we would like to mention, as a theoretical side remark, that the cancellation at $z \to 0$  between the one loop graphs at the threshold for the process $gg \to hh$ is not limited to double Higgs boson production, but also takes place at the thresholds for the processes of $n$-boson production, $gg \to nh$ with {\it even} $n$. It should be mentioned that this behavior is relevant only when $n^2 \mu^2 \ll m^2$ and thus it appears to be not relevant for the actual Higgs boson and the top quark, but can still be useful in other studies. In order to establish this behavior one can use the technique of generating functions for calculating the amplitudes at multiboson thresholds~\cite{brown,agp,mv93}, which automatically takes into account all the tree-type sub-graphs generated by the self interaction in the Higgs sector. Within this approach one calculates the vacuum polarization top quark loop ${\cal A}(k_1,k_2,y)$ in the classical background Higgs field depending on the Euclidean time $\tau$ as
\be
\phi(y)=v+h(y)= v \, { 1+ y/2v \over 1-y/2v}~,
\label{bkgrd}
\ee
where $y=-2 v e^{-\mu \tau}$, so that the field $\phi$ is the familiar solution to the classical equations of motion: $\phi = v \, \tanh(\mu \tau/2)$. Each threshold amplitude $A_n = A(gg \to n h)$ is then found  as the $n$-th derivative with respect to $y$ of ${\cal A}(k_1,k_2,y)$ at $y=0$. In other words the latter is the generating function for all the amplitudes $A_n$ as
\be
{\cal A}(k_1,k_2,y) = \sum_{n=0}^\infty {A_n \over n!} \, y^n~.
\label{gf}
\ee
In the limit, when the ratio $\mu/ m$ is very small, one can consider the variation of the background field on the scale $\mu$ as adiabatic and use the `free' expression for the quark loop with a varying mass $m(y)= m \, [1+h(y)/v]$. The generating amplitude ${\cal A}$, as well as all the amplitudes $A_n$, have the same one form factor structure as in Eq.(\ref{deff}),
\be
{\cal A}(k_1,k_2,y) = -{\alpha_s \over 4 \pi} \, (k_1^\mu \epsilon_1^{a \nu} - k_1^\nu \epsilon_1^{a \mu })(k_{2 \, \mu} \epsilon_{2 \, \nu}^a - k_{2 \, \nu} \epsilon_{2 \, \mu}^a ) \, {\cal F}(y)~,
\label{defcf}
\ee
and one can write for the generating form factor the expression 
\be
{\cal F}={2 \over 3} \, \log \left ( 1+ {h(y) \over v} \right ) = {2 \over 3} \, \log { 1+ y/2v \over 1-y/2v}~,
\label{cfy}
\ee
and, upon the Taylor expansion, the formula for the threshold form factors $F_n$:
\be
F_n= \left [ 1+ (-1)^{n-1} \right] \, {2 \over 3} \, { (n-1)! \over (2 v)^n}~.
\label{fnf}
\ee
Clearly, these form factors are vanishing at even $n$ as a simple consequence of  ${\cal F}$ in Eq.(\ref{cfy}) being an odd function of $y$. It is interesting to note that for odd $n$, where the result in Eq.(\ref{fnf}) is nonzero, there is still a certain cancellation between the (poligon) graphs taking place. Indeed, the contribution to $F_n$ of the triangle graph alone can be evaluated similarly to Eq.(\ref{hexp}):
\be
F_{n (\triangle)} = {2  \over 3} \, \langle nh | \, {h \over v} \, | 0 \rangle = {2  \over 3} \,  {2 \, n! \over (2 v)^n}~,
\label{fn3}
\ee
where the production amplitude $\langle nh | \, h \, | 0 \rangle$ can be found in Ref.~\cite{brown}. The triangle contribution is thus larger than the full result (\ref{fnf}) by the factor $n$: $F_{n (\triangle)}/F_n=n$.

Lacking a full calculation of the amplitude ${\cal A}(k_1,k_2,y)$ beyond the adiabatic in $\mu/m$ approximation, the form factors $F_n$ are not known at arbitrary $z$. We thus can note here only a limited result regarding a generalization of Eq.(\ref{ims}) to $n > 2$. Namely the imaginary part of $F_n(z)$ on the unitary cut at $z > 4/n^2$, associated with the process $gg \to t \bar t \to n h$ with on-shell quarks, is uniquely determined by the zeros~\cite{mv93} of the $t \bar t \to nh$ amplitude as a function of $z$ and the matrix element  $\langle nh | \, h \, | 0 \rangle$ and is given by
\be
{\rm Im}F_{n \, (s)}= {\pi \over 2} \, \theta \left ( z-{4 \over n^2} \right ) \, {2 \, n! \over (2 v)^n} \, z^{-1} \prod_{k=1}^n \left ( 1- {4 \over k^2} \, z^{-1} \right ) \, \log {1+ \sqrt{1-4/(n^2 z)} \over 1- \sqrt{1-4/(n^2 z)}} ~, 
\label{imns}
\ee
although it is not clear at present whether this cut dominates the behavior of the form factor $F_n(z)$ for a general $n$ as it does for $n=2$.

In summary. We have derived the closed analytical expression in Eq.(\ref{fullf}) for the amplitude of $gg \to hh$ at the threshold of the two Higgs bosons, where the cancellation between the tringle and the box graphs of Fig.~1 is the strongest. The reasons for this cancellation are traced to the vanishing of the amplitude as a function of the mass ratio $z=\mu^2/m^2$ at both $z \to 0$ and $z \to \infty$ and, to an extent, to the `extra' zero of the imaginary part of the amplitude on the `major' cut, related to the property of `nullification' of the on-shell threshold amplitudes $t \bar t \to nh$. The strong cancellation between the one loop contributions leads to that the main, in the limit $z \to 0$, two loop correction is numerically comparable to the one loop result for the actual masses of the Higgs boson and the top quark. We have also illustrated that the cancellation in the double boson production amplitude is in fact a part of a more general phenomenon of suppression of multi boson production by two gluons at the corresponding thresholds. 

This work is supported, in part, by the DOE grant DE-FG02-94ER40823.


\begin{thebibliography}{99}
\bibitem{hc} 
  S.~Chatrchyan {\it et al.}  [CMS Collaboration],
  Phys.\ Rev.\ Lett.\  {\bf 110}, 081803 (2013)
  [arXiv:1212.6639 [hep-ex]].
\bibitem{ha} 
  G.~Aad {\it et al.}  [ATLAS Collaboration],
  Phys.\ Lett.\ B {\bf 716}, 1 (2012)
  [arXiv:1207.7214 [hep-ex]].
\bibitem{psz} 
  T.~Plehn, M.~Spira and P.~M.~Zerwas,
  Nucl.\ Phys.\ B {\bf 479}, 46 (1996)
  [Erratum-ibid.\ B {\bf 531}, 655 (1998)]
  [hep-ph/9603205].
\bibitem{cgmppw} 
  R.~Contino, M.~Ghezzi, M.~Moretti, G.~Panico, F.~Piccinini and A.~Wulzer,
  JHEP {\bf 1208}, 154 (2012)
  [arXiv:1205.5444 [hep-ph]].
\bibitem{hkmt} 
  N.~Haba, K.~Kaneta, Y.~Mimura and E.~Tsedenbaljir,
  arXiv:1311.0067 [hep-ph].
  
\bibitem{dkmz} 
  A.~Djouadi, W.~Kilian, M.~Muhlleitner and P.~M.~Zerwas,
  Eur.\ Phys.\ J.\ C {\bf 10}, 45 (1999)
  [hep-ph/9904287].
  
\bibitem{gm} 
  R.~Grober and M.~Muhlleitner,
  JHEP {\bf 1106}, 020 (2011)
  [arXiv:1012.1562 [hep-ph]].

\bibitem{pv} 
  G.~Passarino and M.~J.~G.~Veltman,
  Nucl.\ Phys.\ B {\bf 160}, 151 (1979); \ B.~A.~Kniehl,
  Phys.\ Rev.\ D {\bf 42}, 3100 (1990).
\bibitem{mv93} 
  M.~B.~Voloshin,
  Phys.\ Rev.\ D {\bf 47}, 2573 (1993)
  [hep-ph/9210244].
\bibitem{egn} 
  J.~R.~Ellis, M.~K.~Gaillard and D.~V.~Nanopoulos,
  Nucl.\ Phys.\ B {\bf 106}, 292 (1976).
\bibitem{svvz} 
  M.~A.~Shifman, A.~I.~Vainshtein, M.~B.~Voloshin and V.~I.~Zakharov,
  Sov.\ J.\ Nucl.\ Phys.\  {\bf 30}, 711 (1979)
  [Yad.\ Fiz.\  {\bf 30}, 1368 (1979)].
\bibitem{brown} 
  L.~S.~Brown,
  Phys.\ Rev.\ D {\bf 46}, 4125 (1992)
  [hep-ph/9209203].
\bibitem{agp} 
  E.~N.~Argyres, R.~H.~P.~Kleiss and C.~G.~Papadopoulos,
  Nucl.\ Phys.\ B {\bf 391}, 42 (1993).

\end{thebibliography}
\end{document}